\documentclass[sigconf,authorversion,nonacm]{acmart}

\usepackage{multirow}
\usepackage{xcolor}
\usepackage{microtype}
\usepackage{graphicx}
\usepackage{subfigure}
\usepackage{booktabs} 
\usepackage{natbib} 
\usepackage{hyperref}

\AtBeginDocument{%
  }

\begin{document}

\title{Predicting Customer Lifetime Value Using Recurrent Neural Net}


\author{Huigang Chen}
\affiliation{%
 \country{Meta Platforms, Inc.}
}
\email{huigang@fb.com}

\author{Edwin Ng}\authornote{Contribution was made while the author was at Uber Technologies, Inc.}
\affiliation{%
 \country{Amazon.com, Inc.}
}
\email{edng@amazon.com}

\author{Slawek Smyl}
\affiliation{%
 \country{Meta Platforms, Inc.}
}
\email{slaweks@fb.com}

\author{Gavin Steininger }
\affiliation{%
 \country{Uber Technologies, Inc.}
}
\email{gavin.steininger@uber.com}


\begin{abstract}
This paper introduces a recurrent neural network approach for predicting user lifetime value in Software as a Service (SaaS) applications.
The approach accounts for three connected time dimensions.

These dimensions are the user cohort (the date the user joined), user age-in-system (the time since the user joined the service) and the calendar date the user is an age-in-system (i.e., contemporaneous information).
The recurrent neural networks use a multi-cell architecture, where each cell resembles a long short-term memory neural network. The approach is applied to predicting both acquisition (new users) and rolling (existing user) lifetime values for a variety of time horizons. It is found to significantly improve median absolute percent error versus light gradient boost models and Buy Until You Die models.

\end{abstract}



\keywords{LTV Prediction, RNN, LSTM, GRU}

\maketitle
\section{Introduction}

Services without fixed fees, such as Software as a Service (SaaS) models (e.g., free-to-play gaming), represent a significant global industry. In SaaS, it is crucial to evaluate user value to calculate the return on advertising spend (ROAS) as a predictive function of user acquisition and user engagement. Direct measurement of ROAS, while straightforward, is impractical as it may take months / years to realize the revenue of a user but marketing decisions have to be made sooner. Unlike traditional products with a defined price, in the software as a service model, value is determined by the user, not the business. Thus it is a common practice to predict the value of a user given their early interactions with the service. Creating these predictions can be non-trivial. This work explores a Recurrent Neural Network (RNN) - based approach to predicting user value. 

Predicting user value has two unique challenges. First the user value matures over time. 
That is, if a prediction is needed for total value at age-in-system at each time the prediction must be updated and the uncertainty decreased. 
Second there are three connected time dimensions in addition to user level features. 
These dimensions are the user cohort (the time the user onboarded to the service), user age-in-system (the time since the user joined the service) and the date of the prediction (contemporaneous information; i.e., the state of the service at the time of prediction). 

\section{Problem Statement}
\subsection{Acquisition LTV}

Let $z_t$ be the user value for an arbitrary user for a single age-in-system period $t$ (Note the subscript for users is suppressed for notational clarity). 

The label quantity $Z_{a}^{+}$ is the sum from the current age-in-system period $a$ to the the graduation period $T_0$. That is,

$$Z^{+}_a = \sum_{t=a+1}^{T_0}z_t .$$

Further let 

$$Z^{-}_a = \sum_{t=0}^{a}z_t ,$$

and

$$ Z = Z^{-}_a + Z^{+}_a = \sum_{t=0}^{T_0}z_t .$$ 

In practice, for each individual series, at age-in-system $a$, the historic user values sequence $z_t$, $\forall t \in 1, 2, 
\dots , a$ are observed, while the future values  $z_t$, $\forall t \in  a + 1, a+2, \dots, T_0$  are unobserved.

\subsection{Rolling LTV}

On a given date $t_0$, the goal is to forecast the following quantity at different $T_1, T_2 , \dots, T_k$  for any user

$$Z_{{t_0 + 1}, t_0 + T_j} = \sum^{t_0 + T_j}_{t=t_0 + 1} z_t, 1 \leq{j}\leq{k} $$

Let $t_0$ be the forecast age-in-system, $T_1, \dots, T_k$ as forecast horizons, and $Z_{{t_0 + 1}, t_0 + T_j}$ 
and  as the rolling LTV with horizon $T_j$. Typical forecast horizons include $1$ week, $1$ month, $1$ quarter, $1$ half year etc. 
The forecast may use any information that is available to the forecaster up to the date $t_0$, including $z_{t_0}$, $z_{t_0 - 1}$, and so on.

\begin{figure}
\centering
  \includegraphics[width=0.9\columnwidth]{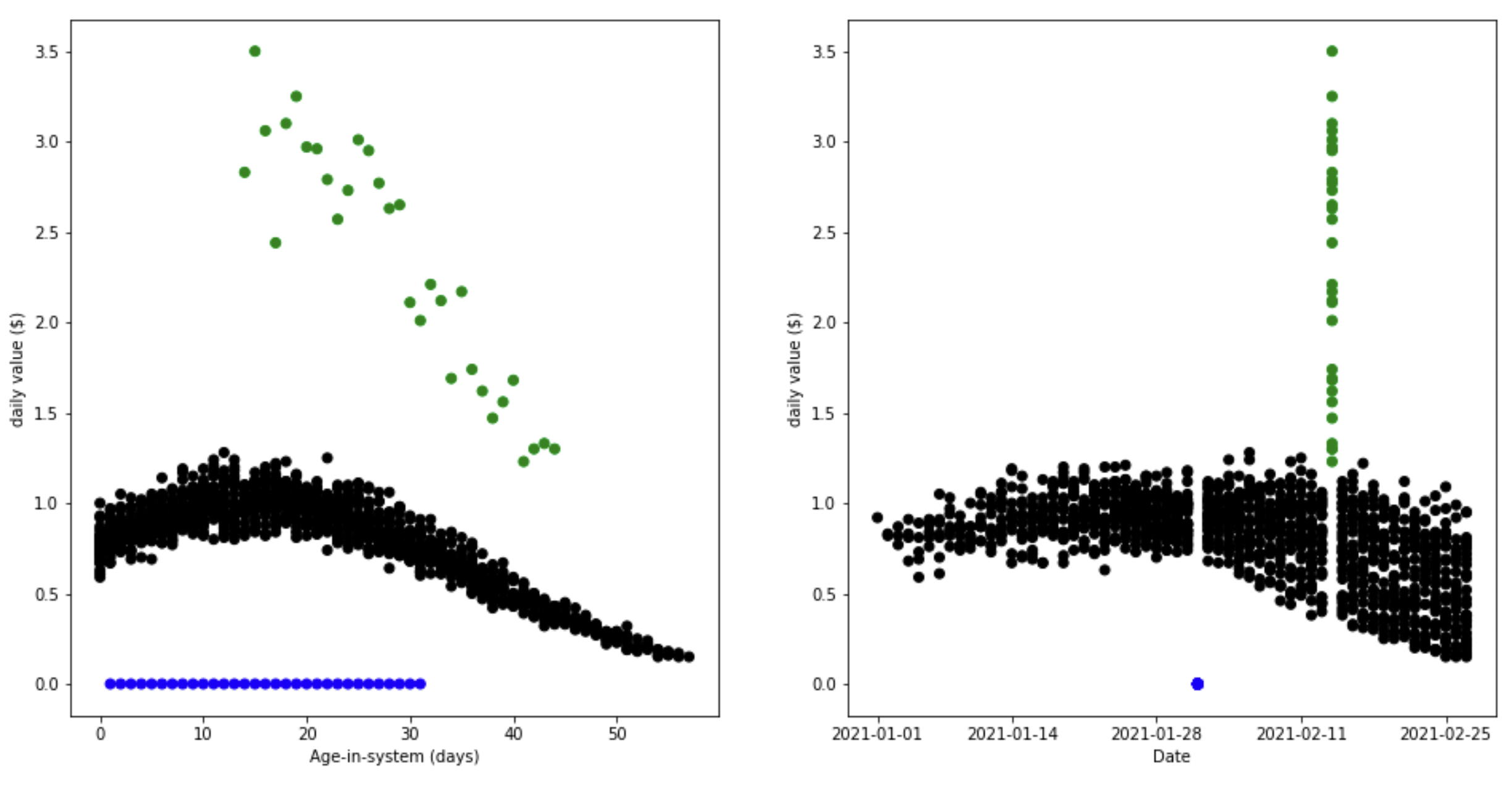}
  \caption{The daily user value as a function of age-in-system (left), and as a function of observation date (right). The colors indicate the system status (blue is an outage, green is an event).}~\label{fig:data-vis}
\end{figure}

While in principle the data can be arranged in a columnar matrix, it is more easily understandable in wide format. 
Table~\ref{table-value-at-age} gives an example of simulated data. 
The diagonal striped pattern highlights that each cohort achieves an age-in-system at a different date. 
Figure ~\ref{fig:data-vis} shows the daily value for an expanded simulated data set both as a function of age-in-system and date. 
The event (green) and outage (blue) data are distributed over age-in-system values but have only one date value each.


\begin{table}
\caption{This table shows daily user value (\$) where row represents specified user and columns represents the age-in-system. Each diagonal with the same text color represents the calendar date when the user reached the given age.}
\label{table-value-at-age}
\begin{flushleft}
\begin{small}
\begin{sc}
\begin{tabular}{| c | c | c | c | c | c | c | c |}

\hline
User ID & Cohort Date & \multicolumn{6}{|c|}{System-in-Age} \\
\hline
& & 0 & 1 & 2 & 3 & 4 & 5  \\
\hline
128 & 2021-01-29 & \textcolor{red}{0.82} & \textcolor{orange}{0.96} & \textcolor{yellow}{0.77 }& \textcolor{brown}{0.00} & \textcolor{teal}{0.78} & \textcolor{cyan}{1.00} \\
\hline
129 & 2021-01-30 & \textcolor{orange}{0.87} & \textcolor{yellow}{0.78} & \textcolor{brown}{0.00} & \textcolor{teal}{0.87} & \textcolor{cyan}{0.99} & \textcolor{blue}{?} \\
\hline
130 & 2021-01-31 & \textcolor{yellow}{0.74} & \textcolor{brown}{0.00} & \textcolor{teal}{0.85} & \textcolor{cyan}{1.03} & \textcolor{blue}{?} & \textcolor{violet}{?} \\
\hline
\end{tabular}
\end{sc}
\end{small}
\end{flushleft}

\vspace{5px}

\begin{flushleft}
\begin{small}
\begin{sc}
\begin{tabular}{| c | c | c | c | c |}
\hline
\multirow{2}{*}{Date} & \textcolor{red}{2021-01-29 }& \textcolor{orange}{2021-01-30} & \textcolor{yellow}{2021-01-31} & 
\textcolor{brown}{2021-02-01}\\
\cline{2-5}
  & \textcolor{teal}{2021-02-02 } & \textcolor{cyan}{2021-02-03}  & \textcolor{blue}{2021-02-04} & \textcolor{violet}{2021-02-05} \\

\hline
\end{tabular}
\end{sc}
\end{small}
\end{flushleft}

\vskip -0.1in
\end{table}

Thus to accurately predict the value of users it necessary to account for both cohort effects (i.e, users that have been onboarded at the same date tend to have similar characteristics, and consequently similar monetization patterns), the contemporaneous effects (the events of a given date / time have an impact for all users active at that time), the age-in-system at which the prediction is to be made, and the potentially user level features (these can and in general do change with age-in-system). It should be noted that while it will likely never be possible to forecast contemporaneous effects (e.g., it is extremely difficult to know that a non scheduled system outage is going to occur at a given time ) it is still necessary to model them to ensure that the cohort and age effects are correctly accounted for.

\section{Related Work}
Traditional approaches to predicting non-contractual user level customer values rely on recency, frequency and monetary value (RFM)\cite{doi:10.1509/jmkr.2005.42.4.415}  from the user’s past history to extrapolate future purchasing behaviors. A prominent model family in this class is a set of parametric generative models aptly named Buy Till You Die (BTYD) \cite{doi:10.1509/jmkr.2005.42.4.415} \cite{doi:10.1287/mksc.1080.0482}. The BTYD approach breaks the customer value forecast into 3 separate modeling objects, the transaction frequency, the transaction amount, and the duration of staying active, and models each independently by different distributions with user level heterogeneous parameters following common prior distributions across all the users, and finally combine them to calculate the forecast value. While the approach is elegant and parsimonious, there are a number of drawbacks that lead to worse performance and limited scope in practical situations. First, the assumption that the transaction amount, the transaction frequency, and the active duration are independent from each other rarely holds in reality. Frequent customers typically are more satisfying customers or have better intent in using the service and they stay with the service longer. Second, it is hard to incorporate useful features, such as seasonality, into a BTYD model, as the model itself is a pure vintage based model. Thirdly, for new users with less transactions, the model does not have enough data to discriminate between potentially high value users and low value ones. Finally, if there are several lines of business,  e.g. Uber Ride and Uber Drive, each has to be modeled independently, with a loss of available  information. Therefore a more expressive model form is often necessary to leverage a wider range of signals available to the forecaster.

Supervised machine learning models that directly use the customer value as the target are the natural candidates to remedy the aforementioned issues with the BTYD models. Popular choices include regularized regression models, random forest and gradient boosting trees. Along with a flexible model form with no distribution assumptions, they need a large amount of features to increase forecast accuracy. On the other hand, these models inherently assume independent observations. Therefore to model the customer value process, a time series object, one needs to heavily rely on feature engineering to capture the temporal dependency. Specifically, one often needs to build out features at different lags explicitly for such models. This may over inflate the feature space and run into runtime performance issues.

\section{Method}

\subsection{The Recurrent Neural Network Approach}

This work proposes the use of a Recurrent Neural Network (RNN) with Long Short-Term Memory (LSTM)-style cells. For the acquisition model, the target variables represent the sum of user daily values (UV) from their current age-in-system to graduation.That is, for every new day, part of the total UV is actualized during the journey of the user from 0 days age-in-system to 90 days age-in-system. Hence, the unknown part that needs to be predicted are the UV “residuals”; i.e., the amount of UVs that are not actualized yet. For the rolling model the target is the sum of a user's next UV for a specified number (varying over use-case) of weeks.

\subsection{Architecture}

\begin{figure}
\centering
  \includegraphics[width=0.9\columnwidth]{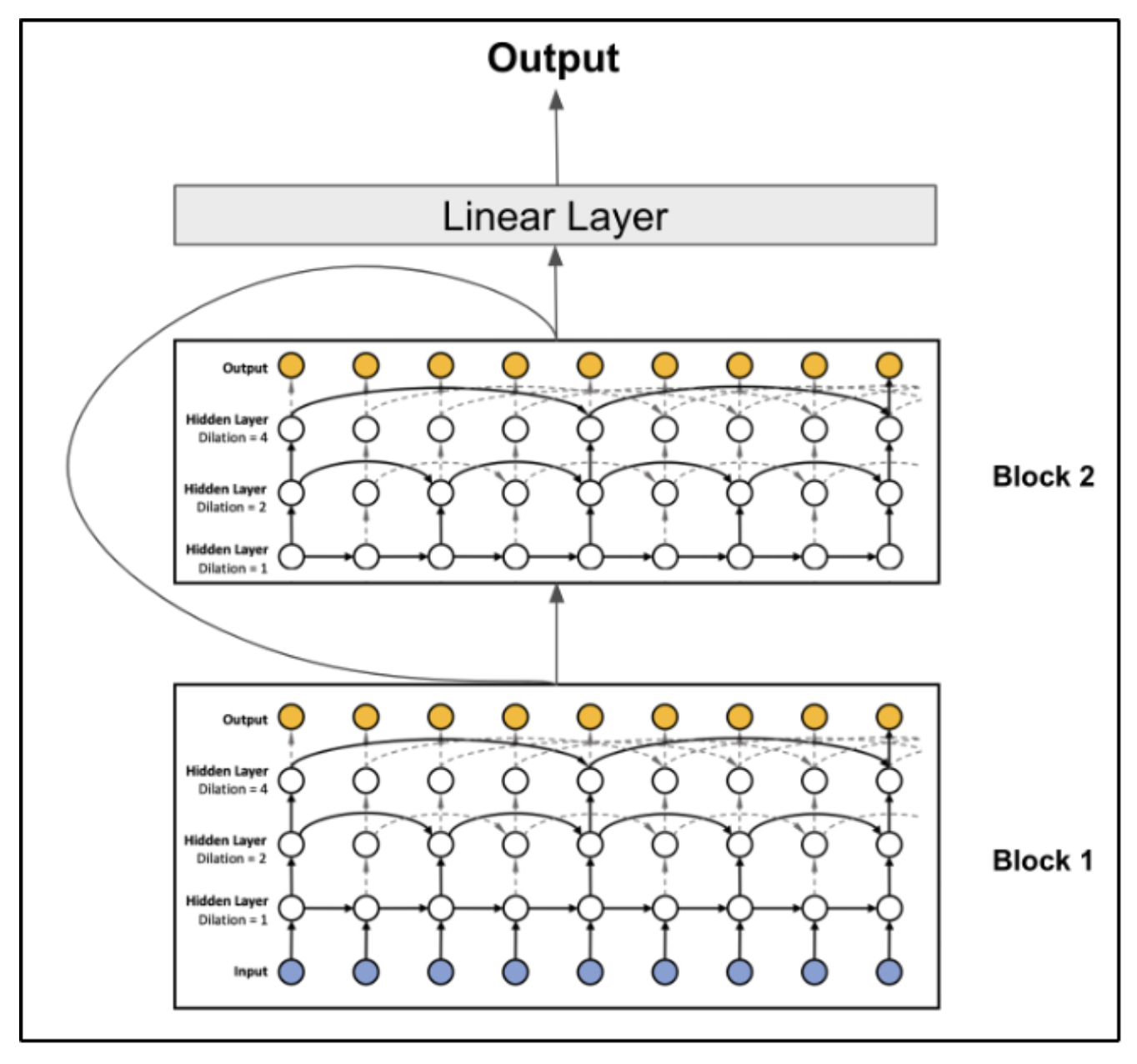}
  \caption{An example RNN composed with 6 layers(cells), in two blocks, with the Res-net style shortcut over block number 2. Additionally, some layers are dilated, and there is a final linear “adaptor” layer that converts the top-most cell output into the needed size. Each dot represents a cell or input or output in a particular time step.}~\label{fig:rnn-architecture}
\end{figure}

Generally, an RNN uses a number of cells, in our case these are standard LSTMs or more advanced LSTM-like cells. It is easy to code an RNN that can plug-in various cells, while keeping the same overall architecture, for example one depicted on Figure~\ref{fig:rnn-architecture}.

It is usually advantageous to use dilated RNNs \cite{NIPS2017_32bb90e8} In those networks, cells use not just the most recent state, but delayed one, by 2 or more steps, like in layers 2, 3, 5, and 6 in Figure~\ref{fig:rnn-architecture}. Finally, a useful trick is to use Res-net like shortcuts \cite{https://doi.org/10.48550/arxiv.1512.03385} between groups of cells, called here blocks.

\subsection{dRNN Cell}

A dRNN is an alternative to the standard LSTM, derived from the LSTM \cite{hochreiter1997long} and GRU \cite{https://doi.org/10.48550/arxiv.1406.1078} cells, and designed to operate as part of a multilayer dilated RNN\cite{NIPS2017_32bb90e8}. 
As in \cite{shwartz2017Sequence} the dRNN output is split into “real output” $Y$, that goes to the next layer, and a controlling output $H$, that is an input to the gating mechanism in following time steps. This last feature (the controlling output) deals with an important limitation of the LSTM, and most other cells: the output at step $t$ is reused at step $t+d$ (where $d>=1$) as a controlling input. 
But there is no reason to assume it is optimal for these two vectors to be the same.

The cell uses two states, c-state (also called cell state), which is close to standard LSTM or GRU state, and $h$-state, which is the controlling state. At each time step, the whole input is a concatenation of $x_t$ , $h_{t-1}$, $h_{t - d}$ , where $x_t$ is a standard input at a time $t$ (either from a previous layer or an input to the RNN), $h_{t-1}$ is the most recent $h$-state, and $h_{t-d}$ is the delayed state ($d>=2$). The cell also uses a fusion gate to create a weighted combination of previous and delayed $c$-states. 
It tends to be more accurate than standard LSTM and GRU cells on forecasting tasks \cite{https://doi.org/10.48550/arxiv.2112.02663}.

\begin{figure}
\centering
  \includegraphics[width=0.9\columnwidth]{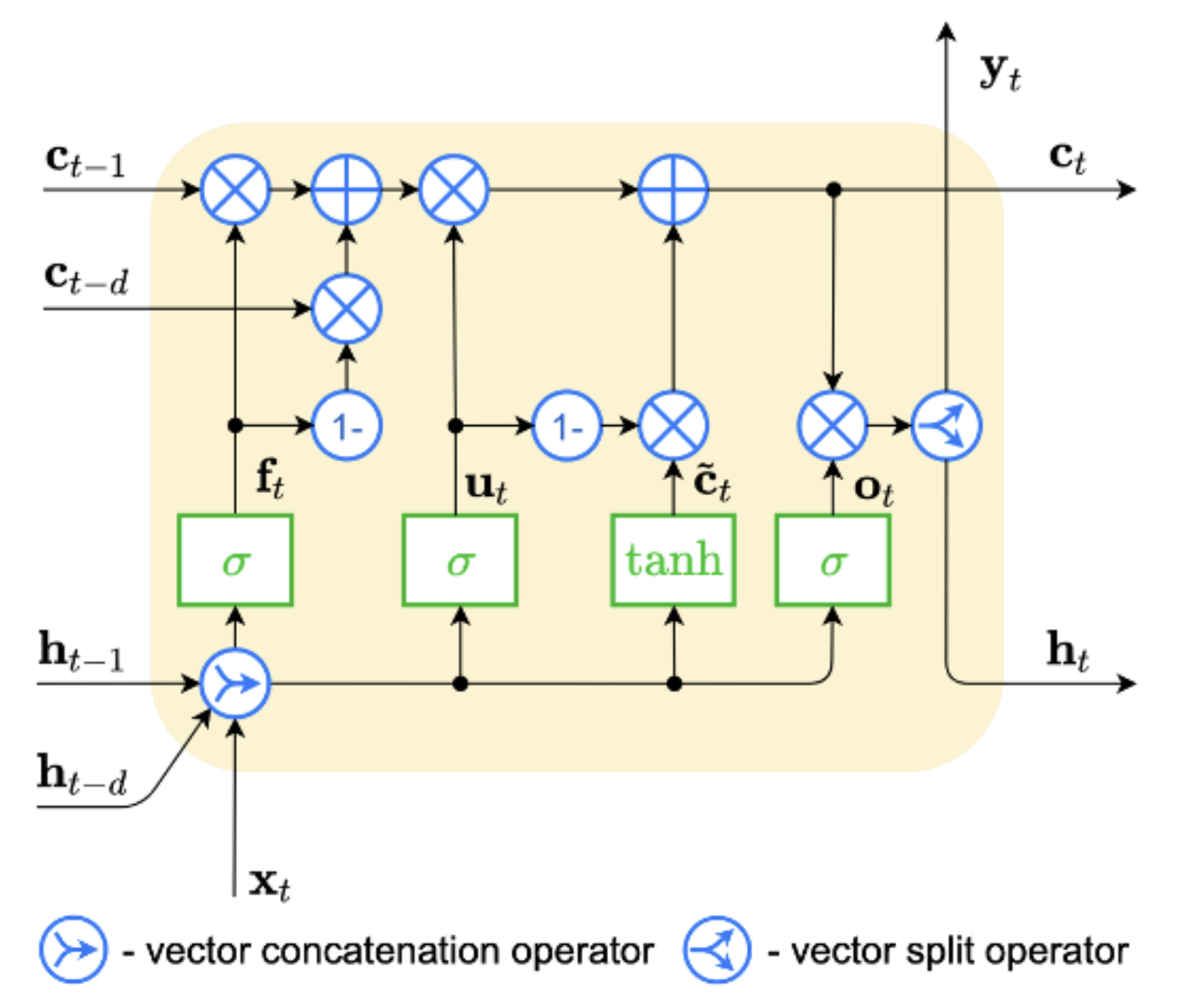}
  \caption{}~\label{fig:dRNN}
\end{figure}

\subsection{Data PreProcessing}

\begin{figure}
\centering
  \includegraphics[width=0.9\columnwidth]{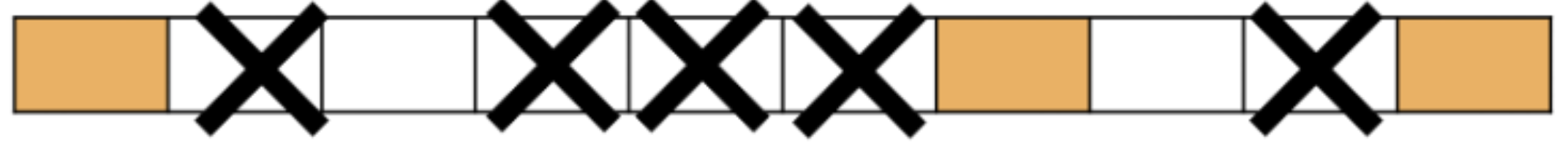}
  \caption{Subsampling of periods with no transactions}~\label{fig:subsampling}
\end{figure}

When forecasting at a user level, apart from standard steps of data preparation, normalization etc., there is also the issue of data sparsity. In particular,  most users purchase infrequently. Looking back one or several weeks, it can be common to find no transactions. The data set is skewed for such “zero-records”. To account for this subsampling of zero-records is used. The process is illustrated in Figure~\ref{fig:subsampling}.

The subsampling causes uneven RNN steps, to compensate for this calendar features, such as week of the year and number of weeks since the last step are added.

\section{Results}

The RNN architecture is applied to two problems: acquisition LTV (used at Uber) and rolling LTV (used at Meta). 
As these problems have different data and use cases different metrics are used to evaluate each case. 
For acquisition LTV root mean squared error is used. 
For rolling LTV a modified symmetric median percent error is used. 
In both cases the RNN models are evaluated against several other models; these models are different for both use cases.

\subsection{Acquisition LTV (Uber)}

The Acquisition LTV was evaluated using root mean squared error (RMSE). 
%
%
%
Two RNN's with 2 LSTM cells are used. First a model that used only convensional features is tested (the model is denoted 2-cell RNN). This model is not found to fit the data as well as an xgboost model. 
Second a 2-cell RNN model was fit with the adition of embedding type features (encodings of cities) is tried (denoted 2-cell embedding). 
The 2-cell embedding model preforms better than the xgboost model. The details are shown in Tab.~\ref{acq-rmse-table}.

\begin{table}[t]
\caption{Summary of models accuracy metric RMSE. Best (the lowest) result is marked in bold.}
\label{acq-rmse-table}
\begin{center}
\begin{small}
\begin{sc}
\begin{tabular}{lcccr} \hline

Model & RMSE \\ \hline

xgboost & 2,292   \\

2-cell RNN  & 2,830\\

2-cell embedding  & \textbf{2,271}  \\ \hline

\end{tabular}
\end{sc}
\end{small}
\end{center}
\vskip -0.1in
\end{table}


\subsection{Rolling LTV (Meta)}
We implement the Rolling LTV forecast model for the Meta Quest users. The Meta Quest is the consumer virtual reality (VR) ecosystem where users make app/in-app purchases from hundreds of apps for a diverse VR experience. We forecast gross revenues from app/in-app purchases at the user level with 1/4/13/26 week forward-looking horizons. The data consists of user level time series up to 4 years long, with 100+ features including demographical, behavioral, and seasonal variables. Feature embedding is also used to reduce the dimension for large categorical variables. 

For Meta Quest user level Rolling LTV forecast, an issue with standard MAPE as the accuracy metric is the prevalence of zero actual values. The relative error size measured by MAPE with zero or close-to-zero actual values is by definition large even for an error of small absolute size. This is not meaningful in most use cases. Therefore to avoid over penalizing small errors when the actual values are zero or close-to-zero, we propose a modification of SMAPE, adjusted SMAPE (aSMAPE), for a given forecast horizon $k$,

$$\text{aSMAPE} = \frac{2}{N} \sum_i \frac{\left| A_{i, k} - F_{i, k} \right|}{\left| \max\{A_{i, k} , a\} + F_{i, k}  \right|},$$
where $N$ is the total number of users in the forecast. Here $a$ is a use case dependent dollar amount where we set a floor on the absolute error for the corresponding relative error metric. For example, if we set $a$ at \$1.00, with 10\% aSMAPE we treat any absolute error under \$0.10 no worse than \$0.10. The use of SMAPE over MAPE has an added benefit that SMAPE is bounded, which is important for a user level accuracy metric.

A comparison of out-of-sample aSMAPE errors between 3 models on a set of Meta Quest users is shown in \autoref{rolling-smape-table}. Here we set $a=\$1.00$. We find that nRNN model consistently outperform BTYD and LGB models.

\begin{table}[t]
\caption{Summary of models accuracy metric aSMAPE in different forecast horizons. Best (the lowest) result is marked in bold.}
\label{rolling-smape-table}
\begin{center}
\begin{small}
\begin{sc}
\begin{tabular}{lcccr} \hline

Model & 4W Forecast & 13W Forecast & 26W Forecast \\ \hline

BTYD & 53.6\% & 70.5\%& 78.3\% \\

LGB &30.4\% & 53.7\% & 61.1\% \\

dRNN & \textbf{22.9}\% & \textbf{39.8}\% & \textbf{51.2}\% \\ \hline

\end{tabular}
\end{sc}
\end{small}
\end{center}
\vskip -0.1in
\end{table}

\section{Conclusion}

The article discusses an RNN approach to model customer lifetime value. It further demonstrates the architecture and technical details on how a variant of neural-net forecasting architecture can be applied in LTV modeling practically. The proposed method provides better forecast accuracy and flexibility, compared to the traditional method such as BTYD and a tree-based approach based on the studies conducted in two types of LTV model from Uber Technology Inc. and Meta. 

\bibliographystyle{SIGCHI-Reference-Format}
\bibliography{ltv-rnn-citation} 

\end{document}